 \documentclass[pdflatex,sn-mathphys-num]{sn-jnl}

\usepackage{graphicx}%
\usepackage{multirow}%
\usepackage{amsmath,amssymb,amsfonts}%
\usepackage{amsthm}%
\usepackage{mathrsfs}%
\usepackage[title]{appendix}%
\usepackage{xcolor}%
\usepackage{textcomp}%
\usepackage{manyfoot}%
\usepackage{booktabs}%
\usepackage{algorithm}%
\usepackage{algorithmicx}%
\usepackage{algpseudocode}%
\usepackage{listings}%
\usepackage{float}
\usepackage{subcaption}
\usepackage{hyperref}
\usepackage{amsmath}
\usepackage{bbold}
\usepackage{graphicx} 
\usepackage{physics}
\usepackage{import}
\usepackage[normalem]{ulem} 
\usepackage{todonotes}
\usepackage[pdftex]{pict2e}
\usepackage[justification=centering]{caption}

\usepackage{tikz}
\usepackage{ifthen,xstring}
\usetikzlibrary{calc}
\usetikzlibrary{external}
\usepackage{graphicx}
\usepackage[dvipsnames]{xcolor}
\usetikzlibrary{hobby}   
\usepackage[outline]{contour}
\usetikzlibrary{fadings}

\usetikzlibrary{decorations.markings}
\usetikzlibrary{arrows.meta}
\usetikzlibrary{arrows}
\usetikzlibrary{decorations}
\usepackage{nicematrix}

\usepackage{tcolorbox}
\tcbuselibrary{breakable}

\newtcolorbox{redbox}{breakable,colback=red!5!white,colframe=red!75!black}
\newtcolorbox{bluebox}{breakable,colback=blue!5!white,colframe=blue!75!black}
\newtcolorbox{greenbox}{breakable,colback=green!5!white,colframe=green!75!black}

\theoremstyle{thmstyleone}%
\newtheorem{theorem}{Theorem}
\newtheorem{proposition}[theorem]{Proposition}%

\theoremstyle{thmstyletwo}%

\theoremstyle{thmstylethree}%
\newtheorem{definition}{Definition}%

\usepackage{etoolbox}
\usepackage{epstopdf}

\begin{document}
\makeatletter
\makeatother
\definecolor{lblue}{rgb}{.1,.6,1}
\colorlet{dyell}{yellow!75!black}
\colorlet{dgreen}{green!50!black}
\colorlet{orng}{yellow!50!red}
\def\isocolor{BurntOrange}
\def\unicolor{Aquamarine}
\def\nodecolor{LimeGreen}

\definecolor{norm}{HTML}{79ADDC}
\definecolor{conj}{HTML}{ffc09f}
\definecolor{data}{HTML}{ffee93}
\definecolor{proj}{HTML}{ffe247}
\definecolor{colora}{HTML}{eba7b1}
\definecolor{colorb}{HTML}{a7d3eb}
\definecolor{greeny}{HTML}{cfdcb6}
\definecolor{purpy}{HTML}{c3b6dd}
\definecolor{gray}{HTML}{c8c6cd}
\def\norm{norm}
\def\conj{conj}
\def\data{data}
\def\proj{proj}
\def\colora{colora}
\def\colorb{colorb}
\def\purpy{purpy}
\def\greeny{greeny}
\def\gray{gray}

\newcommand{\TNode}[4]{ 
	\coordinate (temp) at (#2);
	\ifthenelse{\equal{#1}{}}{
		\draw[line width=1pt,fill=#3] (temp)circle(.25);}
	{\ifthenelse{\equal{#1}{C}}{
			\draw[rounded corners=2pt,line width=1pt,fill=#3] ($(temp)-(.25,.25)$)rectangle++(.5,.5);}
		{\ifthenelse{\isodd{#1}}{
				\draw[rounded corners=2pt,line width=1pt,fill=#3] (temp)--++($(.25,0)+#1*(0,.25)$)--++($#1*(0,-.5)$)--++(-.5,0)--(temp);
            \coordinate (temp) at ($(temp)-#1*(0,.05)-(.05,0)$);}
			{\draw[rounded corners=2pt,line width=1pt,fill=#3] (temp)--++($(-.25,0)+#1*(0,.125)$)--++($#1*(0,-.25)$)--++(.5,0)--(temp);
            \coordinate (temp) at ($(temp)-#1*(0,.05)-(.05,0)$);}}}
	\node at (temp) {#4}}

\newcommand{\Gate}[5]{ 
	\coordinate (temp) at (#2);
	\ifthenelse{\equal{#1}{V}}{
		\draw[rounded corners=2pt,line width=1pt,fill=#4!50!white] ($(temp)-(.25,.25)$)rectangle++($(-.25,.5)+#3*(.75,0)$)node[midway]{#5}}
	{\draw[rounded corners=2pt,line width=1pt,fill=#4!50!white] ($(temp)-(.25,-.25)$)rectangle++($(.5,.25)-#3*(0,.75)$)node[midway]{#5}}}

\newcommand{\Qin}[2]{ 
	\coordinate (temp) at (#2);
	\StrLen {#1}[\tmpLn]
	\ifthenelse{ \tmpLn= 1}{
	\ifthenelse{\equal{#1}{V}}{
	\draw[line width=1pt, fill=yellow!75!red] ($(temp)-(0,.1)$)circle(.15)}
	{\draw[line width=1pt, fill=yellow!75!red] ($(temp)-(.1,0)$)circle(.15)}}
	{\ifthenelse{\equal{#1}{V0}}{
		\node[below] at (temp) {$|0\rangle$}}
		{\node[left,align=right] at (temp){$|0\rangle$}}}}

\newcommand{\Qout}[3]{ 
	\coordinate (temp) at (#2);
	\StrLen {#1}[\tmpLn]
	\ifthenelse{ \tmpLn= 1}{
	\ifthenelse{\equal{#1}{V}}{
		\draw[line width=1pt] ($(temp)+(-.1,.05)$) --++(.2,.1)node[above]{#3};
		\draw[line width=1pt] ($(temp)+(-.1,0)$) --++(.2,.1)}
	{\draw[line width=1pt] ($(temp)+(-.05,.1)$) --++(.1,-.2)node[right]{#3};
	\draw[line width=1pt] ($(temp)+(0,.1)$) --++(.1,-.2)}}{
	\ifthenelse{\equal{#1}{VM}}{\coordinate (temp2) at ($(temp)+(0,.25)$);}{\coordinate (temp2) at ($(temp)+(.25,0)$);}
	\draw[line width=1pt] ($(temp2)-(.25,.25)$)rectangle++(.5,.5);\draw[line width=1pt,->] ($(temp2)+(-.25,-.25)$)--++(.35,.35);}}

\newcommand{\Qreset}[3]{
	\coordinate (temp) at (#2);
	\ifthenelse{\equal{#1}{V}}{
		\ifthenelse{\equal{#3}{in}}{
			\draw[fill=white,color=white] ($(temp)-(.1,-.1)$)rectangle++(.2,.4);
			\Qout{V}{$(temp)$}{};
			\Qin{V}{$(temp)+(0,.5)$}}{
			\ifthenelse{\equal{#3}{out}}{\draw[fill=white,color=white] ($(temp)-(.1,-.1)$)rectangle++(.2,1.2);
				\Qout{VM}{$(temp)$}{};
				\Qin{V0}{$(temp)+(0,1.2)$}}{\draw[fill=white,color=white] ($(temp)-(.1,-.1)$)rectangle++(.2,.4);
				\Qout{V}{$(temp)$}{};
				\Qin{V0}{$(temp)+(0,.5)$}}}}{
		\ifthenelse{\equal{#3}{in}}{
			\draw[fill=white,color=white] ($(temp)-(0,.1)$)rectangle++(.4,.2);
			\Qout{H}{$(temp)$}{};
			\Qin{H}{$(temp)+(.5,0)$}}{
			\ifthenelse{\equal{#3}{out}}{
			\draw[fill=white,color=white] ($(temp)-(0,.1)$)rectangle++(1.1,.2);
			\Qout{HM}{$(temp)$}{};
			\Qin{H0}{$(temp)+(1.2,0)$}}{\draw[fill=white,color=white] ($(temp)-(0,.1)$)rectangle++(.65,.2);
			\Qout{H}{$(temp)$}{};
			\Qin{H0}{$(temp)+(.75,0)$}}}}}

\def\ctrlsize{.1}

\newcommand{\MPSiso}[4]{
    \coordinate (I1) at #1;
	\ifthenelse{\equal{#4}{d}}
	    {
	        \ifthenelse{\equal{#3}{r}}
	        {   
	            \draw[rounded corners=2pt,line width=1pt,fill=#2]  
                (I1)--++(.25,.25)--++(-.5,0)--++(0,-.5)--(I1)}
	        {
	            \draw[rounded corners=2pt,line width=1pt,fill=#2] 
                (I1)--++(-.25,.25)--++(.5,0)--++(0,-.5)--(I1)}}
	    {   
	        \ifthenelse{\equal{#3}{r}}
	        {
                \draw[rounded corners=2pt,line width=1pt,fill=#2] 
                (I1)--++(-.25,.25)--++(0,-.5)--++(.5,0)--(I1)}
	        {   
                \draw[rounded corners=2pt,line width=1pt,fill=#2] 
                (I1)--++(.25,.25)--++(0,-.5)--++(-.5,0)--(I1)}}        
}

\newcommand{\Projector}[2]{
    \coordinate (I1) at (#1);
    \draw[rounded corners=2pt,line width=1pt,fill=#2]  
                (I1)--++(.25,0)--++(0,-.25)--++(-.5,0)--++(0,.25)--(I1);
}

\newcommand{\symiso}[2]{
    \coordinate (I1) at (#1);  
    \draw[line width=1pt,fill=#2] 
     (I1)--++(0.35,0)--++(-0.35,0.35)--++(-0.35,-0.35)--(I1);
}

\newcommand{\symuni}[2]{
    \coordinate (I1) at (#1);
    \draw[line width=1pt,fill=#2] 
     ($(I1)$)--++(0.25,0)--++(0,0.5)--++(-.5,0)--++(0,-.5)--($(I1)$);
}

\newcommand{\symnode}[2]{
    \coordinate (I1) at ($(#1)$);
    \draw[line width = 1pt, fill=#2] (I1) circle (.25);
}

\newcommand{\Cnot}[4]{
    \def\circlesize{#3}
    \def\ctrlsize{#4}
    \coordinate (temp) at (#1);
    \draw[line width = 1pt] (temp)--++($(0,#2)$);
    \draw[fill=black] (temp) circle (\ctrlsize);
    \draw[fill=white,line width = 1pt] ($(temp)+(0,#2)$) circle (\circlesize);
    \draw[line width = 1pt] ($(temp)-(\circlesize,-#2)$) --++ ($(2*\circlesize,0)$);
    \draw[line width = 1pt] ($(temp)-(0,-#2+\circlesize)$) --++ ($(0,2*\circlesize)$);
    }
    
\newcommand{\Cphase}[2]{
    \coordinate (temp) at (#1);
    \draw[line width = 1pt] (temp)--++($(0,#2)$);
    \draw[fill=black] (temp) circle (\ctrlsize);
    \draw[fill=black] ($(temp)+(0,#2)$) circle (\ctrlsize);
    }

\newcommand{\Cunitary}[4]{
    \coordinate (temp) at (#1);
    \draw[line width = 1pt] (temp)--++($(0,#2)$);
    \draw[fill=black] (temp) circle (0.2*#4);
    \draw[line width = 1pt, fill = white,rounded corners = 3] ($(temp)+(-#4,-#4+#2)$) rectangle ++($(#4*2,#4*2)$);
    \node[line width = 1pt] at ($(temp)+(0,#2)$) {#3};
    }

\newcommand{\Hadamard}[1]{
    \coordinate (temp) at (#1);
    \draw[line width = 1pt, fill = white] ($(temp)+(-0.1,-0.1)$) rectangle ++(0.2,0.2);
    \node[line width = 1pt] at (temp) {H};
    }
\newcommand{\unitaryone}[2]{
    \coordinate (temp) at (#1);
    \draw[line width = 1pt, fill = white] ($(temp)+(-0.1,-0.1)$) rectangle ++(0.2,0.2);
    \node[line width = 1pt] at (temp) {#2};
    }
    
\newcommand{\unitary}[4]{
    \coordinate (temp) at (#1);
    \draw[line width = 1pt, fill = white, rounded corners = #4] ($(temp)+(-.5*#3,-.5*#3)$) rectangle ++(#3,#3);
    \node at (temp) {#2};
    }
\newcommand{\readout}[4]{
    \coordinate (temp) at (#1);
    \draw[line width = 1pt] ($(temp)+(.05,-.15*#2)$) --++(0,#4);
    \draw[line width = 1pt] ($(temp)+(-.05,-.15*#2)$) --++(0,#4);
    \draw[line width = 1pt,fill=black] ($(temp)+(0,-.15*#2+#4-.4*#2)$) --++(.2*#2,.4*#2)--++(-.4*#2,0) -- cycle;
    \draw[line width = 2pt, fill = white, rounded corners = #3] ($(temp)+(-.5*#2,-.5*#2)$) rectangle ++(#2,#2);
    \begin{scope}
        \clip  ($(temp)+(-.4*#2,-.15*#2)$) rectangle ++(.8*#2,.6*#2);
        \draw[line width = 2pt] ($(temp)+(0,-.15*#2)$) circle(.3*#2);
    \end{scope}
    \draw[line width = 2pt] ($(temp)+(0,-.15*#2)$) --++(.3*#2,.4*#2);
}
    
\newcommand{\unitarytwo}[3]{
    \coordinate (temp) at (#1);
    \draw[line width = 1pt, fill = white] ($(temp)+(-0.1,-0.1)$) rectangle ++($(0.2,0.2+#2)$);
    \node[line width = 1pt] at ($(temp)+(0,#2*0.5)$) {#3};
    }

\title[Article Title]{Entanglement is Half the Story: Post-Selection vs. Partial Traces}

\author*[1]{\fnm{Gustav J L} \sur{Jäger}\orcid{https://orcid.org/0000-0002-4002-5259}}\email{gustav.jaeger@dlr.de}

\author[1]{\fnm{Krzysztof} \sur{Bieniasz}\orcid{https://orcid.org/0000-0003-2583-4338}}

\author[2]{\fnm{Martin B} \sur{Plenio}\orcid{https://orcid.org/0000-0003-4238-8843}}

\author[1]{\fnm{Hans-Martin} \sur{Rieser}\orcid{https://orcid.org/0000-0002-1921-1436}}

\affil[1]{\orgdiv{Institut für KI-Sicherheit}, \orgname{Deutsches Zentrum für Luft- und Raumfahrt e.V.}, \orgaddress{\street{Wilhelm-Runge-Straße 10}, \city{Ulm}, \postcode{89081}, \country{Germany}}}

\affil[2]{\orgdiv{Institut für Theoretische Physik and IQST}, \orgname{Universität Ulm}, \orgaddress{\street{Albert-Einstein-Allee 11}, \city{Ulm}, \postcode{89081}, \country{Germany}}}


\abstract{
While tensor networks have their traditional application in simulating quantum systems, in the recent decade they have gathered interest as machine learning models. We combine the experience from both fields and derive how quantum constraints placed on a tensor network manifest a change in capabilities. To this end, we employ a method of inference of classical tensor networks on a quantum computer to define a hybrid architecture. This hybrid tensor network is a practical unified framework for it’s classical and quantum tensor network edge cases. We identify post-selection as the important property on which this interpolation hinges. The amount of post-selection corresponds to the level to which quantum constraints are enforced on the tensor network. On this basis, we propose a new hyperparameter which controls the transition between the hybrid and the quantum tensor network. In the comparison of classical and quantum tensor networks it complements the bond dimension. Quantum machine learning is improved by using the hyperparameter to allocate the practically limited post-selection to the quantum model in a trainable manner.
}




\maketitle
\section{Introduction}\label{sec: introduction}
With machine learning (ML) becoming a vital field of the sciences, the field of quantum computing has spawned the interest in quantum machine learning (QML). In ML, non-linearities are vital for attaining good results, e.g. prominent activation functions like the sigmoid in NN \citep{Grohs2022}. On the other hand, QML struggles with imparting non-linearities due to the constraints on quantum systems and their evolution \citep{Schuld2021}. However, quantum effects like superposition and entanglement as well as the exponentially growing Hilbert space may enable efficient solutions to classically hard problems \citep{Nielsen2012, Schuld2021}. In this work, we will take a step towards the understanding between classical and quantum machine learning by addressing and outlining the differences between the ML architectures of classical and quantum tensor networks (TNs). We outline how these differences can be quantified, proposing a hyperparameter to the TN architecture in addition to the commonly used bond dimension. We identify the amount of post-selection as the core difference between the two architectures. Furthermore, on this basis we present a method to improve the performance of QML methods in a multi-shot context. It takes advantage of post-selecting ancilla qubits to introduce local non-linearities. To this end, let us introduce TNs with their background in quantum physics and their use case as an ML model.

To begin with, let us look at TNs in their traditional quantum background: Over the last decades TNs have emerged and been developed for the description and simulation of quantum systems \citep{White1992, Ran2020, Bridgeman2017, Orus2013, Schollwoeck2010} and pose a state of the art method for this application \citep{Pan2021, Tindall2023, Berezutskii2025}. Tensor networks compress the description of a quantum system by representing subsystems and their interactions with tensors and their bonds, respectively \citep{Bridgeman2017}. As TNs are classical constructs, they do not necessarily obey the fundamental quantum properties of normalization of the state, hermitian observables, and completely positive trace preserving (CPTP) maps \citep{Bengtsson2017, Wilde2016}. Any TN that globally corresponds to these constraints is called a quantum tensor network (QTN) \citep{Rieser2023}. 
For TNs without loops, these constraints are commonly implemented using orthogonality centers \citep{Orus2013}. Such methods also exist for more complex TNs but they restrict the space of available states \citep{Zaletel2020}. Common QTNs include matrix product states (MPS), tree tensor networks (TTN), multi-scale entanglement renormalization ansatz (MERA), and projected entangled pair states (PEPS) \citep{Ran2020, Bridgeman2017, Orus2013, Schollwoeck2010}. In the context of the young field of QML, QTNs are used as ML models \citep{Huggins2019, Stoudenmire2018}, architectures for variational circuits \citep{Huggins2019}, encodings and classical dimensionality reduction techniques \citep{Dilip2022, Hickmann2025}, and initializations via classical pre-training for quantum models \citep{Wall2022}.

TNs have been applied as classical ML models during the last decade. Dropping quantum constraints, the first architectures were simple methods based on classical tensor networks (CTNs) like MPS, also referred to as tensor trains (TTs) in this context \citep{Stoudenmire2016, Novikov2016}.  
This has been developed further to more complex architectures \citep{Stoudenmire2018, Huggins2019}. Deeper investigations into properties of TNs in the ML context lead to a significant improvement in understanding the combination of both methods, e.g. the impact of mutual information for ordering inputs \citep{Harada2025a}, and observed grokking \citep{Pomarico2025, Pomarico2025a}. Recent efforts seek to embed TNs into larger classical ML setups \citep{Tomut2024, Wang2023}.

Recently, hybrid quantum-classical TNs (HTNs) have been proposed to combine advantages of TN both in quantum and ML\citep{Yuan2021}. One main idea of HTNs is to project the capabilities of quantum computing to a large dimension which currently is not reached by quantum computers or at which quantum computers are still inefficient \citep{Yuan2021, Chen2021, Huang2023, Xiu2023, Schuhmacher2025, Harada2025, Hickmann2025}. In \citep{Yuan2021} more efficient simulations of larger quantum systems were achieved by combining CTNs with a smaller QTN. Others expand the limits of NISQ devices for QML tasks by using a CTN \citep{Huang2023, Harada2025}. One motivation is to increase the size of datasets that can be handled by the hybrid algorithm and compressing the data for QML using classical dimensionality reduction, see \citep{Chen2021, Hickmann2025}. In the end, TNs are envisioned to seamlessly merge ML capabilities of quantum computing with classical approaches \citep{Yao2024}.

The structure of the work is as follows: In Sec.~\ref{ssec: post-selection} we give background on the lack of non-linearities in quantum systems and the need to introduce them as pre- or post-processing steps. In this context, we provide a toy example to demonstrate how post-selection can be used as a post-processing step. In Sec.~\ref{ssec: transformations} we analyze the differences in structure of CTNs and QTNs and propose an inference strategy for CTNs on quantum computers based on post-selection. Sec.~\ref{ssec: hybrid architecture} features the definition of a hybrid tensor network architecture that includes both CTNs and QTNs. Post-selection is identified as the core difference between CTNs and QTNs in this hybrid architecture. Sec.~\ref{sec: model} outlines some practical choices and assumptions including about the classification problem, the normalization of the output, and the loss function. In Sec.\ref{sec: Hyperparameters} we identify the hyperparameter and give the required proof. The numerical results are given in sec.~\ref{sec: num results}. We investigate the behaviour of the hybrid model dependent on the hyperparameter. Lastly, in Sec.~\ref{sec: conclusion} we conclude the work and give an outlook.

\section{Methods}\label{sec: methods}
\subsection{Post-Processing by Post-Selection}\label{ssec: post-selection}

One of the limitations of QML is the lack of global non-linearities. With their linear operations, QML models cannot improve global separation of the data, which is needed for classification tasks. Quantum channels can only compress already existing global separation into a smaller dimensional Hilbert space and thus make it accessible for readout, i.e. they address local separation \citep{Schuld2021}. For improving global separation, QML models need to rely on pre- and post-processing, because these can introduce non-linearities. 
One common approach is the so-called data reuploading, i.e. the sequential multiple encoding of the same classical data into a quantum circuit. This allows for the introduction of non-linearities at the pre-processing level \citep{Schuld2021, Cao2017}. However, in the case of quantum data generated by another quantum system the ability to pre-process is limited. In such cases post-processing is needed. In this section, we identify a viable approach to cause global separation based on post-selection. To this end, let us first give more background on quantum operations.

\medskip\noindent\textbf{Separation of States with Quantum Channels}\\
Commonly in quantum computing, quantum states are represented by normalized complex vectors $\ket{\Psi}$, also known as ket vectors. They are evolved through unitary operators $U$ that perform a rotation in a complex vector space. However, more generally quantum states are represented by positive semidefinite hermitian density matrices $\rho$ with unit trace. The general quantum operations are given by completely positive trace preserving (CPTP) linear maps \citep{Bengtsson2017, Wilde2016}: Any quantum channel $\Lambda$ applied to a mixed quantum state $\rho$ may be expressed as a set of Kraus operators $K_i$, s.t.
\begin{equation}\label{eq: kraus operators}
    \Lambda[\rho] = \sum_i^r K_i \rho K_i^\dag, \quad \text{with} \quad \sum_i^r K_i^\dag K_i = I.
\end{equation}
The smallest number of Kraus operators needed to describe a channel $\Lambda$ is called its Kraus rank $r$ \citep{Bengtsson2017}. Thus, for $r=1$ the Kraus operators have to be isometries that project the quantum state to a larger or equal space but they can not reduce the Hilbert space, i.e. project the Hilbert space to a smaller one. The maximum rank necessary to obtain any channel acting on $\rho$ is $\dim(\rho)\dim(\Lambda[\rho])$ \citep{Bengtsson2017}. Alternatively, a channel can be rewritten using Stinespring's dilation \citep{Bengtsson2017, Wilde2016, Bridgeman2017} to be expressed as a unitary acting on the system $A$ and a pure ancilla system $B$ as
\begin{equation}\label{eq: stinespring dilation}
    \Lambda[\rho_A] = \tr_B\big(U_{AB} (\rho_A \otimes \ketbra{0}_B)U_{AB}^\dag\big).
\end{equation}
From this, it follows that with sufficiently large ancilla system $B$, it is possible to represent a quantum channel acting on $A$ as a linear operation on the larger system. Such a linear operation is not able to globally separate the quantum states for a classification problem any more than they already are. This is caused by the quantum data processing inequality \citep{Wilde2016} 
\begin{equation}\label{eq: data processing inequality}
    S(\Lambda[\rho]\,||\,\Lambda[\sigma]) \leq S(\rho || \sigma)
\end{equation}
which states that any quantum channel may only decrease the relative entropy between two quantum states. A similar statement can be formulated for the cross-entropy. The overall quantum channel may only project the separating information to a smaller subsystem, while the best possible classification is still limited by the encoding and the resulting globally optimal separation. On the full system $AB$ including the ancilla, there exists an optimal split of the Hilbert space into $l$ orthonormal subspaces for a classification of $l$ labels. Then each of these subspaces may be mapped to the computational states $\ketbra{l}$. This quantum map can then be represented in the two manners shown above in Eqs.~\eqref{eq: kraus operators}~and~\eqref{eq: stinespring dilation}. 

There are two ways to further improve the classification result: Either improve the encoding through classical pre-processing or use classical post-processing. The pre-processing path is directly linked to the question: Which problems are well suited for QML and how do we map them to QML? Improving the encoding leads to better results for one-shot quantum circuits as they make the problem easier for the quantum computer. For the post-processing path the question is: How can we use the obtained results efficiently? This question only arises in a multi-shot context, where a statistical ensemble of the output of the quantum computer is aggregated. It is this latter question we aim to answer by comparing QTNs and CTNs.

Let's take a look at an example, how we might improve the performance through post-processing. A simple post-processing step that is common in quantum computing is post-selection, which is simply disregarding certain measurement outcomes from the statistic. This is commonly done without loss of generality by disregarding all the shots of the quantum computer for which the post-selected qubits are not measured as $\ketbra{0}$. Thus, it requires multiple runs on the quantum device to achieve success.

\begin{figure}[H]
    \centering
    \includegraphics{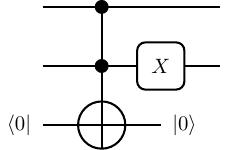}
    \caption{Toffoli gate with post-selection that either separates $\ket{+1}$ and $\ket{1+}$ with certainty or disqualifies the shot with post-selection. After post-selection $\ket{+1}$ is output as $\ket{00}$ and $\ket{1+}$ as $\ket{11}$.}
    \label{fig: qml example}
\end{figure}

\medskip\noindent\textbf{Example -- Post-Selection and Separation of Quantum States}\\
In this example, post-selection addresses the core problem in quantum computing that for the separation of two states $\ket{a}$ and $\ket{b}$ by a circuit, the limit for the separation is given by the initial overlap $|\braket{a}{b}|^2$. In Fig.~\ref{fig: qml example} we choose $\ket{a}=\ket{+1}$ and $\ket{b}=\ket{1+}$. There are no completely orthonormal subspaces to separate our two states because the states have a non-zero overlap. However, the subspace on which the two states overlap can be post-selected out by removing all shots that originate from a $\ket{11}$ input.  

As a result, we obtain that a high level of required post-selection indicates that either the dataset as a whole may be unsuitable to QML because they contain a lot of overlapping information, the circuit ansatz is not suitable, or the encoding translates an insufficient amount of separation between the Hilbert subspaces associated with the different classes, i.e. the problem was not already solved by the classical preprocessing and the QML model would simply project down the information from the inaccessibly large Hilbert space. In the case of directly obtained quantum data without a choice of encoding, post-selection can improve the classification accuracy in a multi-shot environment substantially by identifying and discarding runs that contain overlap between classes. A clear benefit is that the model removes potential misclassifications by post-selecting them out.

\subsection{Moving between Classical and Quantum Tensor Networks}\label{ssec: transformations}
In this section we review the core differences of the architectures of CTNs and QTNs with respect to their application in ML and QML. The fundamental difference is that QTNs equate to a quantum channel and thus any QTN is a CPTP map, see Eqs.~\eqref{eq: kraus operators}~and~\eqref{eq: stinespring dilation}. Additionally, quantum systems can be represented by QTNs with large bond dimensions for which it is unfeasible to contract or even store them classically, due to the exponentially growing Hilbert space. Thus, we observe two qualities that have opposite effects on the expressivity of QTNs as compared to CTNs. While QTNs can benefit from quantum simulation by overcoming the limitation of the bond dimension inherent to all TNs, their expressivity is nonetheless limited by the quantum constraints. There has been a lot of work to address optimization given such constraints, with prominent methods derived from the field of Riemannian optimization \citep{Novikov2016, Boumal2023}. However, in the context of machine learning it is an open question how the introduction of these constraints affects the learning capabilities, which is precisely the question we wish to study here, in order to to gain insight into the potential of QML.

\begin{figure}[H]
    \centering
    \includegraphics{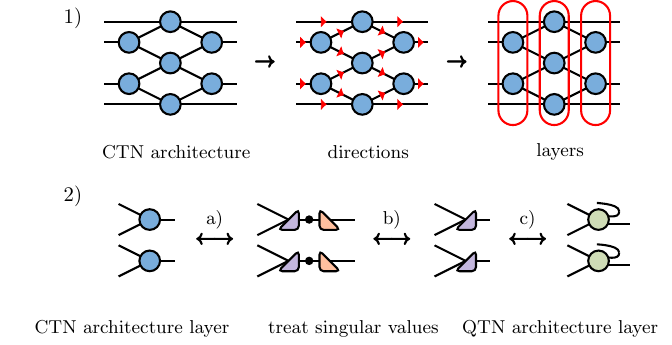}
    \caption{Process for the transition from a CTN architecture to a QTN architecture. 1) General approach of defining the layers is given, which can be applied to both QTNs and CTNs. 2) Transformation of the tensors in a layer necessary for the transition from either CTN to QTN or vice versa. The individual steps are: a) SVD or contraction, b) impose or lift restrictions on singular values, and c) convert between isometries (with post-selection) and unitaries (with partial traces). }
    \label{fig: TN transition}
\end{figure}

\medskip\noindent\textbf{Current Mapping between CTNs and QTNs}\\
The common approach to map between QTNs and CTNs architectures is sketched in Fig.~\ref{fig: TN transition}, see also \citep{Wall2022,Huggins2019}. Quantum constraints are commonly enforced by allowing partial traces and letting each tensor be isometric, s.t. a smaller input is mapped to a larger output according to the directions in Fig.~\ref{fig: TN transition}. Thus, there are no isometries that project a larger input to a smaller output as that would require post-selection. Then, we may convert any CTN to a QTN by restricting the respective singular values to one and converting isometries that project down into unitaries that are partially traced. Any QTN architecture may be converted to a CTN by lifting this restriction on the singular values and turning partial traces into post-selection. In this process new degrees of freedom emerge in either direction, one being the singular values and the other presenting when converting the isometries into unitaries with partial traces.

\begin{figure}[H]
    \centering
    \includegraphics{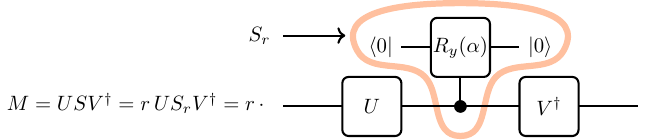}
    \caption{Inference of an arbitrary $2\times2$ matrix on a single qubit through post-selection. The singular values are rescaled by the largest singular value $r$, such that the largest entry is $1$. The other singular values $d=\cos(\alpha/2)$ is implemented using a controlled rotation with angle $a$.}
    \label{fig: matrix decomposition}
\end{figure}

\medskip\noindent\textbf{Inference of CTNs on Quantum Computers}\\
However, QTNs and CTNs can both be executed on universal quantum computers if post-selection is allowed. Generally for any TN, the translation into a quantum circuit requires the grouping of the open indices into inputs and outputs, like the directions in Fig.~\ref{fig: TN transition}. Thus, the TN can be reshaped into a matrix operation. This matrix operation is then formatted to conform to quantum constraints and subsequently transpiled into a quantum circuit. For QTNs this procedure is trivial. CTNs however, require post-selecting and rescaling of the quantum circuit results. This process is demonstrated in Fig.~\ref{fig: matrix decomposition} for a $2\times 2$ matrix. Similar methods are implicitly found in quantum computing approaches to partial differential equations \citep{Siegl2025, Termanova2024} or quantum signal processing \citep{Suzuki2025}, where non-unitary linear maps are applied on quantum computers \citep{Zylberman2025}. The approach from Fig.~\ref{fig: matrix decomposition} can easily be extended to larger matrices by including multi-controlled rotations (MCRY gates). Each MCRY then addresses one of the singular values obtained by the singular value decomposition of the original matrix operation, see Fig.~\ref{fig: matrix decomposition}. In a qubit efficient approach allowing mid-circuit measurements this process requires only a single ancilla qubit. Furhtermore, the approach can also be easily implemented using deferred measurements on the ancilla. Although the required ancilla count then jumps to at most as many ancillas as singular values.

Thus, one can theoretically run inference of any tensor network on a quantum computer, i.e. any QTN or CTN. However, a practical limitation is quickly found in the amount of post-selection that is possible on a quantum computer. Quantum resources are costly, and the number of singular values and thus multi-controlled rotations grow exponentially with the number of qubits. However, more efficient approaches may be derived from common implementations of diagonal unitaries \citep{Zylberman2025}. Practically, the amount of possible post-selection may be given by the number of total available shots and a minimal number of shots necessary for obtaining a statistically meaningful result. This lower bound may be linked to the expected fidelity of the output.

\subsection{Tensor Network Architectures: Classical vs. Quantum}\label{ssec: hybrid architecture}

Given the comparison of QTNs and CTNs in Sec.~\ref{ssec: transformations}, we observe that the core differences lie in bond dimensions, the singular values, and how the Hilbert space is reduced (partial trace or post-selection). Furthermore, we have also seen the protocol that allows for the inference of CTNs on quantum computers and thus their representation as QTNs. Based on these observations we now define a common architecture that allows for a smooth transition between CTNs and QTNs.


\medskip\noindent\textbf{Defining a Hybrid Tensor Network}\\
To this end, we have to address some fundamental choices so that QTNs and CTNs can be compared. From Eqs.~\eqref{eq: kraus operators}~and~\eqref{eq: stinespring dilation} and the correspondence of a QTN to CPTP map, it follows that the QTN can be represented as a TN that is contracted with its complex conjugate. On the other hand, the first CTNs to be used in ML were of an MPS like architecture \citep{Stoudenmire2016, Novikov2016} and contained only real valued tensors. Therefore, to allow a genuine comparison, we widen the CTNs to also contain complex values and similarly allow the contraction with it's complex conjugate. Complex valued classical ML models are not unprecedented given the field of CVNNs \citep{Hirose2012,Lee2022,Barrachina2023}. Furthermore, because QTNs are maps it is valid to assume a grouping into input legs, i.e. bonds, and output legs analogous to the grouping in Fig.~\ref{fig: TN transition}.

\begin{figure}[H]
    \centering
    \includegraphics{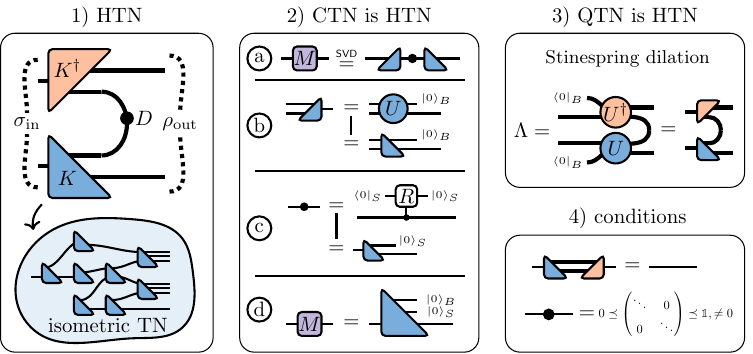}
    \caption{Overview of the proposed HTN with 1) its structure, 2a)-d) rewriting of a CTN into a HTN format, 3) rewriting of a QTN into HTN format, and 4) TN conventions of triangles for isometries and dots for diagonal matrices. }
    \label{fig: HTN explanation}
\end{figure}
We can now start by introducing the HTN and later defining QTNs and CTNs as special cases. 
\begin{definition}\label{prop: HTN}
    Any TN is an HTN if it can be split into three parts: A TN where each tensor is isometric, its conjugate, and a set of real diagonal matrices $0 \preceq D_j \preceq I$ with $D_j \neq 0$, which we will call the reduction operators, which connect the first TN with its conjugate. The isometric TN may only project the input to a larger or equal Hilbert space, i.e. the direction in which it is isometric is relevant. Projection from a larger to a smaller Hilbert space is solely achieved via the reduction operators. With out loss of generality we will aggregate all reduction operators into one $D = \otimes_j D_j$ via tensor product. The parts of the HTN are portrayed in Fig.~\ref{fig: HTN explanation}.1.
\end{definition}
    The outlined architecture in Fig.~\ref{fig: HTN explanation}.1 is able to represent both QTNs and CTNs, see Props.~\ref{prop: QTN}~and~\ref{prop: CTN}. Thus, it serves as a well suited definition for the hybrid of QTNs and CTNs. However, it differs from the understanding of HTNs in previous works, see \citep{Yuan2021, Chen2021, Huang2023, Xiu2023, Schuhmacher2025, Harada2025, Hickmann2025, Yao2024}. While they use a combined architecture whose parts are individually either QTNs or CTNs, we define our HTNs as an architecture that can both take on the role of a CTN or QTN depending on the task. This definition has the core aim of highlighting the differences between CTNs and QTNs.
    
\begin{proposition}\label{prop: QTN}
    QTNs are HTNs with identities as diagonal tensors, i.e., any QTN can be represented as an HTN with all the reduction operators as identities, and any HTN with identities can be viewed as a QTN.
\end{proposition}
    Proof: Using the Stinespring dilation Eq.~\eqref{eq: stinespring dilation} any quantum channel may be represented by a unitary acting on the input and an ancilla system $\ketbra{0}_B$ which is traced out at the end. Absorbing, i.e. contracting, the ancilla into this unitary yields the isometry required for the HTN, see Fig.~\ref{fig: HTN explanation}.3. Then, we identify the partial trace lead to reduction operators $D$ that are identities.
    Thus, any quantum channel can be rewritten in accordance with our definition of an HTN. Furthermore, an isometry followed by partial traces is a CPTP map, as is evident from the Stinespring dilation. Thus, any HTN with identities for the reduction operators is always a QTN.

\begin{proposition}\label{prop: CTN}
    CTNs are HTNs up to a global scaling factor if the reduction operators are one-hot matrices, i.e. they are zero except one diagonal element that is set to one. (This comes with the caveat that only the diagonal entries of the output density matrix are evaluated, which is equivalent to a measurement in the computational basis. This is because CTNs are usually not contracted with their complex conjugate.)
\end{proposition}

Proof: The arguments layed out here are visualized in the TN language in Fig.~\ref{fig: HTN explanation}.2. Firstly, we may take the singular value decomposition of all tensors in the CTN between the bonds in the input and output direction, see Fig.~\ref{fig: HTN explanation}.2a. To this end, each bond has to first be assigned this direction, which might be obtained from the contraction order, see Fig.~\ref{fig: TN transition}. This results in two isometries and a tensor for the singular values in between. Any isometry with a larger input dimension compared to output dimension is isometric in the wrong direction, i.e. it projects down from a larger space to a smaller Hilbert space. The goal is to isolate the Hilbert space reducing operations and implement them via the reduction operators $D$. To this end, any such isometries can be rewritten by a unitary which is contracted by a vector $\ket{\varphi}_B$ often chosen to be $\ket{0}_B$, see. Fig.~\ref{fig: HTN explanation}.2b. This process is commonly known as post-selection in quantum computing. The obtained reduction operators $D_j = \ketbra{0}$ are one-hot. 

All that is left to address are the singular values. They may be implemented using the strategy identified in Sec.~\ref{ssec: transformations}. To this end, they are rescaled s.t. any non-zero diagonal matrix of singular values has it's largest entry as $1$, which results in the global scaling factor mentioned in Prop.~\ref{prop: CTN}. We disregard the global scaling factor because we will later assume a normalization of the output of the HTN, see Sec.~\ref{sec: model}. With a controlled rotation on ancilla qubits and subsequent post-selection of these ancilla qubits, one can implement the singular values using a unitary operation followed by a post-selection, see Fig.~\ref{fig: HTN explanation}.2c. Again the resulting reduction operators are one-hot. Subsequently, any CTN can be written, although cumbersome, as an HTN, see Fig.~\ref{fig: HTN explanation}.2d. 

\begin{figure}[H]
    \centering
    \includegraphics{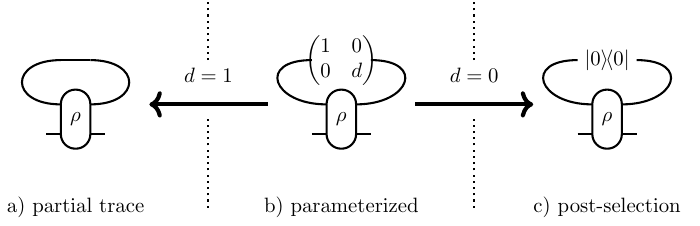}
    \caption{Different approaches to reduce the density operator $\rho$ of a quantum state to a Hilbert space of lower dimension. The parameterized approach in b) for a $2\times 2$ reduction operator has two edge cases: a) the partial trace with $d=1$, and c) post-selection with $d=0$. }
    \label{fig: rank regularization}
\end{figure}

\medskip\noindent\textbf{Parameterized HTN and CTN/QTN Edge Cases}\\
Similar ideas of parameterized diagonal matrices are found in rank regularization for tensor networks \citep{Barratt2022}. However, here they are solely placed on the Hilbert space reducing bonds as reduction operators. Fig.~\ref{fig: rank regularization} portrays how the reduction operators have the edge cases post-selection and partial trace that are associated with CTNs and QTNs respectively.

\section{A HTN ML Model}\label{sec: model}
In Sec.~\ref{ssec: post-selection} we saw an example of how post-selection can be applied as a post-processing step in QML of a classification task. In this section, we introduce a mathematical description of the HTN model similar to CTN machine learning models \citep{Stoudenmire2016}. It can then be used to define and explain the hyperparameters in Sec.~\ref{sec: Hyperparameters}. Additionally to build a complete ML pipeline, we address the loss function and encoding strategy of classical data.

To this end, let's move from a tensor network to a general description of the map that corresponds to an HTN. It can be given as
\begin{equation}\label{eq: post-selected channel}
    \Lambda[\sigma_i]_A=\tr_B(U\sigma_i U^\dag (D_B \otimes I_A)).
\end{equation}
where the $U$ corresponds to the isometric TN found in Figs~\ref{fig: HTN explanation}.1. We choose $U$ to be unitary by moving the ancilla system necessary for isometries $U_{\text{iso}} = U (\ket{0}\otimes I)$ into the encoded data $\sigma_i$. Thus, from here on out $\sigma_i$ contains not only the data but also a sufficiently large system at $\ket{0}$. $D_B$ is the aggregate tensor product of the reduction operators. Overall, the system is divided into subsystems $A$ and $B$, where $A$ is the system of the output and on $B$ is completely removed by the reduction operators.

We apply this map to a classification task and assume that the labels are encoded as pure target states,
\begin{equation}
    \tau_i = \ketbra{\tau_i}{\tau_i}.
\end{equation}
Without loss of generality, we can identify these pure states as the computational basis with the label $l$ encoded as $\ketbra{l}{l}$. Due to the post-selection, the trace of the density operator output by the channel in Eq.~\eqref{eq: post-selected channel} is effectively no longer equal to $1$. To compare the output state with the target states, an amplification of the trace is necessary. In CTNs, it is common to find the diagonal matrices obtained by SVD to have entries larger than 1, which corresponds to a simple multiplying factor equally applied to each output state. However instead of treating each state equally, better losses are achieved if each output state is individually normalized. To this end, we divide each output state $\rho_i=\Lambda[\sigma_i]$ by it's trace,
\begin{equation}\label{eq: normalization}
    \mathcal{N}[\rho_i] = \frac{\rho_i}{\tr (\rho_i)}.
\end{equation}
This step also allows us to disregard the global scaling factors of CTNs.

For non invertible matrices like density operators of pure quantum states, the matrix logarithm in Eq.~\eqref{eq: cross entropy loss} is ill-defined. To circumvent this issue, we use a depolarizing channel $\Delta_\lambda$. The quantum state is set to the maximally mixed state with a small probability $\lambda$ that we choose sufficiently large to ensure numerical stability. The depolarizing channel \citep{Wilde2016} for a quantum state $\rho$ with dimension $d$ is defined as
\begin{equation}\label{eq: depolarizing channel}
    \Delta_\lambda [\rho] = (1-\lambda) \rho + \lambda \frac{I}{d}.
\end{equation}
With the cross entropy loss \citep{Shangnan2021,Grohs2022} we can now obtain our loss function,
\begin{equation}\label{eq: cross entropy loss}
L = \frac{1}{N}\sum_{i=1}^N  -\tr \big(\tau_i \log \Delta_\lambda \circ  \mathcal{N} \circ \Lambda[\sigma_i]\big).
\end{equation}
Similarly to CTN ML models \citep{Stoudenmire2016} the MSE loss might be a more practical choice.

\section{Proposed Hyperparameter}\label{sec: Hyperparameters}
One prominent property of TNs is the bond dimension and in the context of machine learning it is an obvious hyperparameter. Because it can be adapted during training \citep{Stoudenmire2016,Geng2022}, the hyperparamter is typically the maximum bond dimension to which larger bonds are truncated. For CTNs and generally for QTNs the bond dimension is monotonous in the optimized loss, such that a larger bond dimension results in a better or equal loss. However, it is limited classically by storage constraints, while quantum systems have states with arbitrary large bond dimension. Due to the current state of hardware, we use QTNs as a stand in for quantum systems and are thus also limited in the bond dimension.  In a sense the bond dimension is a resource which favours quantum systems. However, given a fixed bond dimension, previous results have shown that the introduced CPTP constraints can significantly impact the performance negatively \citep{Jaeger2025}. This gap in the comparison of CTNs and QTNs is at the core of this work. To this end, we introduce a new hyperparameter, that controls this introduction of quantum constraints on the previously defined HTN. At its heart, this hyperparameter allows for the transition between strict enforcement and complete relaxation of the quantum constraints on the hybrid architecture of the HTN. However, it does not allow for the full control between QTNs and CTNs directly, which would require an additional hyperparameter possibly linked to the purity of the output. Instead it allows us to ascertain how well the QTN performs in the comparison to the overarching HTN architecture. 

Given Eq.~\eqref{eq: cross entropy loss} and $\lambda = 0$, we want to define an $h$-dependent loss function $L_h$, such that
\begin{align}\label{eq: hparam edge losses}
\begin{split}
    L_0 &= \frac{1}{N}\sum_{i=1}^N  -\tr \big(\tau_i \log \Delta_\lambda \circ  \mathcal{N} \circ \Lambda[\sigma_i]\big), \quad \text{and}\\ 
    L_1 &= \frac{1}{N}\sum_{i=1}^N  -\tr \big(\tau_i \log  \Delta_\lambda \circ \Lambda[\sigma_i]\big).
\end{split}
\end{align}
where $h \in [0,1]$ controls whether the normalization is applied or not.
A practical choice is to introduce a threshold $t \in [0,1]$ on how much normalization is applied
\begin{equation}
    \mathcal{N}_t[\rho_i] = \frac{\rho_i}{\max (\tr(\rho_i),t)}.
\end{equation}
This practical approach is oriented towards limiting the percentage of discarded shots. Alternatively, the hyperparameter $w \in [0,1]$ can be motivated from information theory and limits the normalization to
\begin{equation}
    \mathcal{N}_w[\rho_i] = \frac{\rho_i}{\tr(\rho_i)^{(1-w)}}.
\end{equation}
For $\lambda = 0$, i.e. no depolarizing channel, the information weighting emerges by using the properties of the logarithm on $L_w$:
\begin{align}
\begin{split}
    L_w &=  \frac{1}{N}\sum_{i=1}^N  -\tr \big(\tau_i \log \Lambda[\sigma_i]\big) - (1-w) \frac{1}{N}\sum_{i=1}^N  -\log \tr\big(\Lambda[\sigma_i]\big).
\end{split}
\end{align}
We observe that for $L_{h=1}$ the normalization disappears. We show that $L_1$ is minimized by setting all reduction operators equal to identities. Thus, $L_1$ is minimized by avoiding post-selection completely and using partial traces instead. This equates to strictly adhering to the trace preserving part of the CPTP constraint and thus it results in quantum channels, recall Fig.\ref{fig: HTN explanation}. Because $h$ transitions between $L_0$ and $L_1$ it is a suitable hyperparameter to control the permitted amount of post-selection. 

While for $L_0$ post-selection is allowed, it is not enforced in the strictest sense as $L_0$ is not necessarily minimized by setting all reduction operators to one-hot diagonal matrices. Thus, the hyperparameter transitions between a QTN and HTN, but not a CTN. 

\begin{proposition}\label{prop: no norm}
    Given the loss $L_1$ in Eq.~\eqref{eq: hparam edge losses}, without affecting the minimized loss one can assume that all diagonal matrices of the HTN architecture in Fig.~\ref{fig: QML model} can be set to the identity.
\end{proposition}

Proof: For a single instance of the dataset $i$ with the $\sigma_i$ as the encoded data combined with a sufficiently large ancilla system, we may write the output $\rho_{i,D}$ of the HTN according to Eq.~\eqref{eq: post-selected channel} as
\begin{equation}
    \rho_{i,D} = \tr_B \Big(U\sigma_iU^\dag(D_B\otimes I)\Big),
\end{equation}
while the QTN output is
\begin{equation}
    \rho_{i,I} = \tr_B \Big(U\sigma_iU^\dag \Big).
\end{equation}
Then
\begin{equation}
    \rho_{i,I} - \rho_{i,D} = \tr_B \Big[U\sigma_iU^\dag((I-D_B)\otimes I)\Big] \succeq 0.
\end{equation}
Since the depolarizing channel preserves order and the logarithm is operator monotone on positive semidefinite matrices \citep{Chansangiam2015,HIAI2010,Furuta2013},
\begin{equation}
    -\tr[\tau_i \log \Delta_\lambda(\rho_{i,I})] \leq - \tr[\tau_i \log \Delta_\lambda(\rho_{i,D})].
\end{equation}
Thus for every (U,D), the choice $D=I$ is no worse, proving the optimum of $L_1$ can be taken at the QTN edge. The proof for the MSE loss is found in App.~\ref{app: MSE loss proofs}. 

Furthermore, we can make the following argument:
\begin{proposition}\label{prop: monotonicity}
The losses $L_w$ and $L_t$ change monotonously with the hyperparameters $w$ and $t$ respectively.
\end{proposition}
Proof: For $\mathcal{N}_t$
\begin{equation}
    t_1 < t_2 \Rightarrow \max(\tr[\rho_{i,D}],t_1) \leq \max(\tr[\rho_{i,D}],t_2)\Rightarrow
    \mathcal{N}_{t_1} \succeq \mathcal{N}_{t_2}.
\end{equation}
Again invoking the order preserving of the depolarizing channel and the operator monotone property of the logarithm, we obtain
\begin{equation}
    -\tr \big(\tau_i \log \Delta_\lambda \circ \mathcal{N}_{t_1} \circ \Lambda[\sigma_i]\big) \leq -\tr \big(\tau_i \log \Delta_\lambda \circ \mathcal{N}_{t_2} \circ \Lambda[\sigma_i]\big).
\end{equation}
Thus, the loss $L_t$ is monotone with the hyperparameter $t$. Similarly for $\mathcal{N}_w$
\begin{equation}
    w_1 < w_2 \Rightarrow (\tr[\rho_{i,D}])^{w_1} > (\tr[\rho_{i,D}])^{w_2} \Rightarrow \mathcal{N}_{w_1} \succeq \mathcal{N}_{w_2}.
\end{equation}
Like the above result, $L_w$ is monotone with $w$.
\section{Numerical Results}\label{sec: num results}
\begin{figure}[H]
    \centering
    \includegraphics{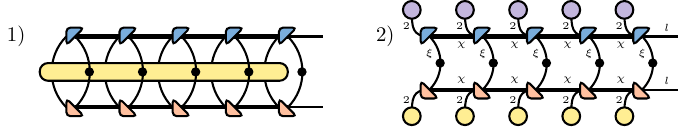}
    \caption{ML model for an MPS-inspired HTN. 1) is assuming an arbitrary encoding of the data into the yellow MPO. 2) show the rotational encoding into the purple and yellow product states. The blue and orange isometric tensors are their respective complex conjugates at each site. The black dots are the reduction operators, i.e. real diagonal matrices $0\preceq D\preceq I$ with $D\neq 0$ and dimension $\xi$. Contracting all bonds leaves the output state on the open legs on the left.}
    \label{fig: QML model}
\end{figure}

To investigate the HTN from Def.~\ref{prop: HTN} we utilize a simple MPS based architecture. While more complex TN architectures offer a wider range of applications, MPS is still a common choice for investigating new TN methods or modifications of the architecture. The MPS based HTN is portrayed in Fig.~\ref{fig: QML model}.1. It allows us to investigate the effect of post-selection on our system. 
For the numerical experiments, we choose rotational encoding, see Fig.~\ref{fig: QML model}.2. The complexity of the HTN is overall quartic in the bond dimension, while classical MPS inspired ML models are quadratic, see \citep{Stoudenmire2016}. The product state encoding can help address scaling issues, which is why they are commonly employed for CTN based ML models. In this section, we aim to demonstrate how the previous findings might be applied. Because we focus specifically on the post-processing step, we do not survey other encoding schemes. Similar to \citep{Stoudenmire2016} we use environment caching during the optimization sweeps, which for the HTN scales quadratic with the bond dimension of the channel, the bond dimension of the encoding, and linear with the size of training data. or a more detailed guide on environment caching refer to \citep{Le2025}. Furthermore, like \citep{Stoudenmire2016} we use an initialization of the tensor network using the data. We tested the initializations stability and for the extreme cases of $\chi$, $\xi$, and $t$. We found the loss of our initialization to be within 4$\%$ of the smallest loss of 200 random intializations and within 3$\%$ of their average.

\begin{figure}[H]
    \centering
    \begin{subfigure}{0.49\textwidth}
        \includegraphics[width=\textwidth]{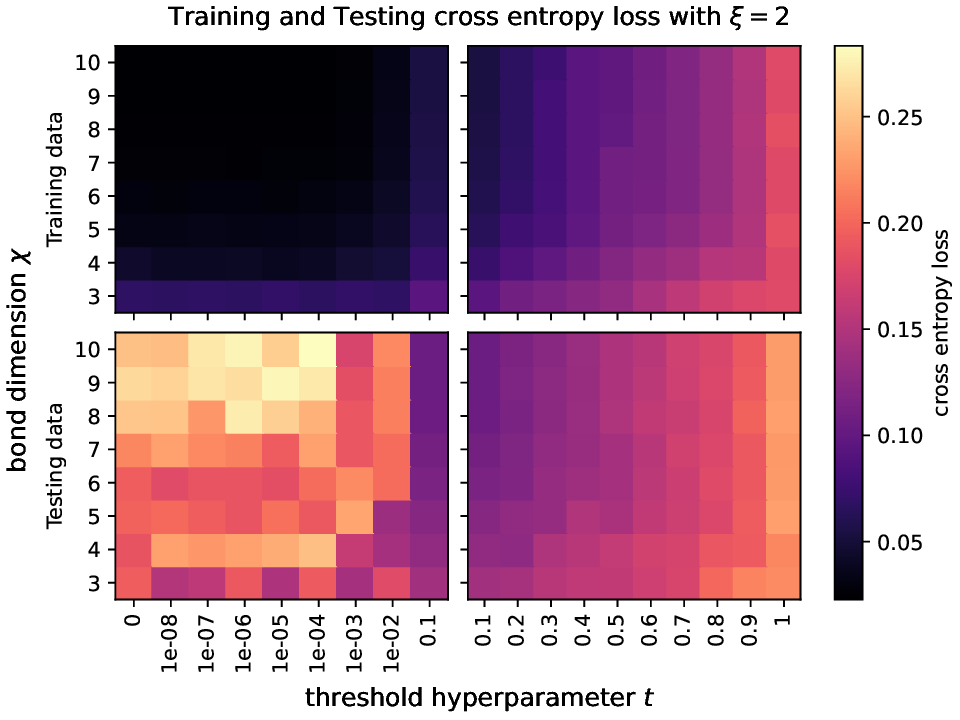}
        \caption{Loss for $\xi = 2$}
        \label{fig: xi2 loss}
    \end{subfigure}
    \begin{subfigure}{0.49\textwidth}
        \includegraphics[width=\textwidth]{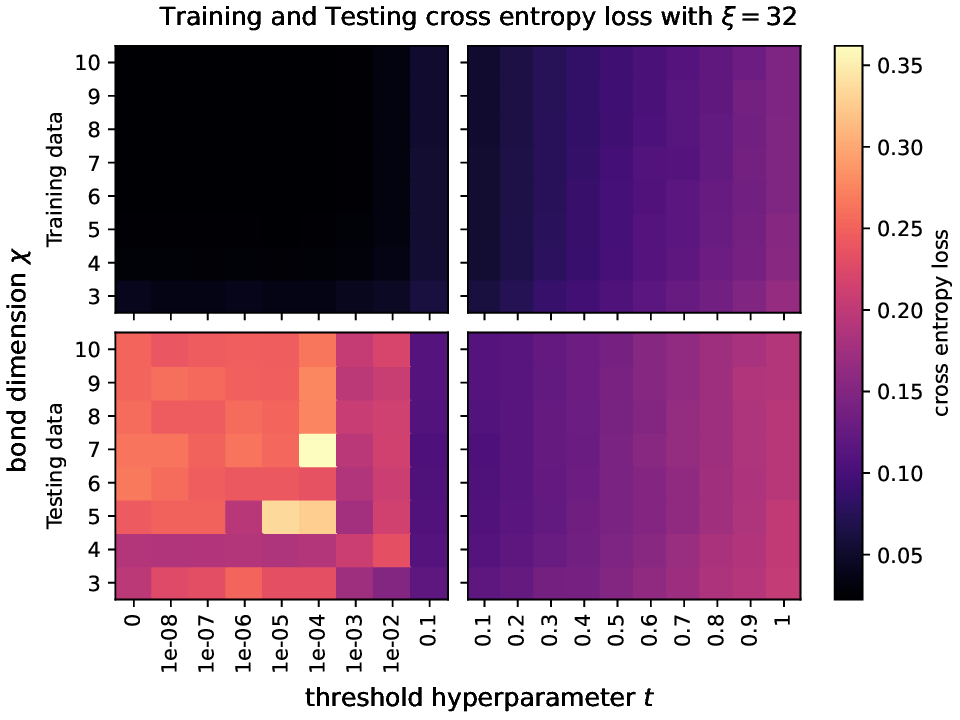}
        \caption{Loss for $\xi = 32$}
        \label{fig: xi32 loss}
    \end{subfigure}
    \begin{subfigure}{0.49\textwidth}
        \includegraphics[width=\textwidth]{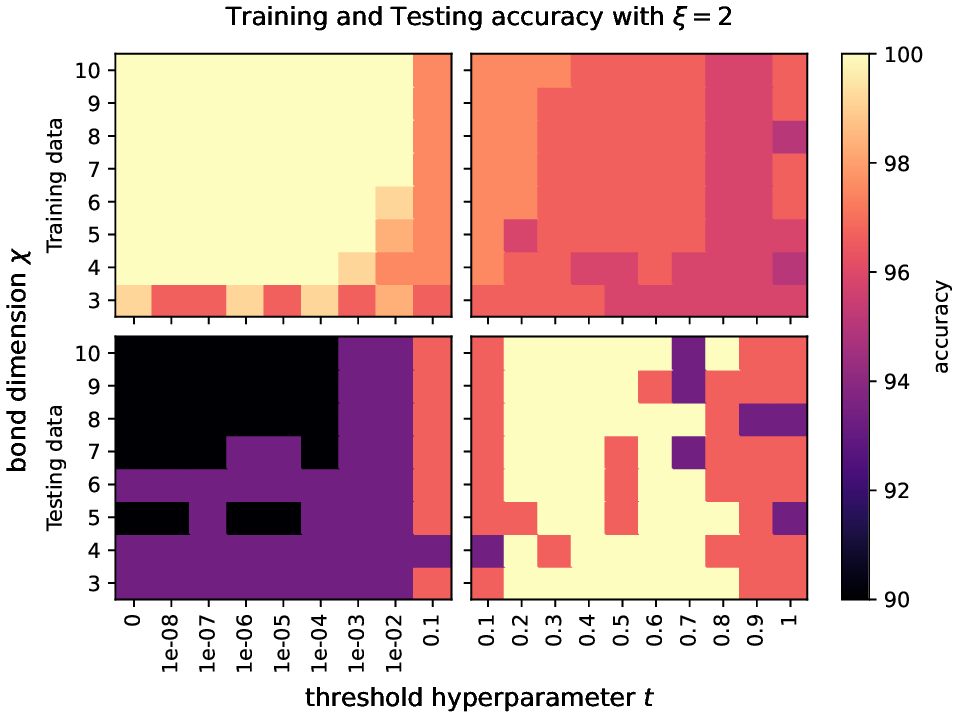}
        \caption{Accuracy for $\xi = 2$}
        \label{fig: xi2 accuracy}
    \end{subfigure}
    \begin{subfigure}{0.49\textwidth}
        \includegraphics[width=\textwidth]{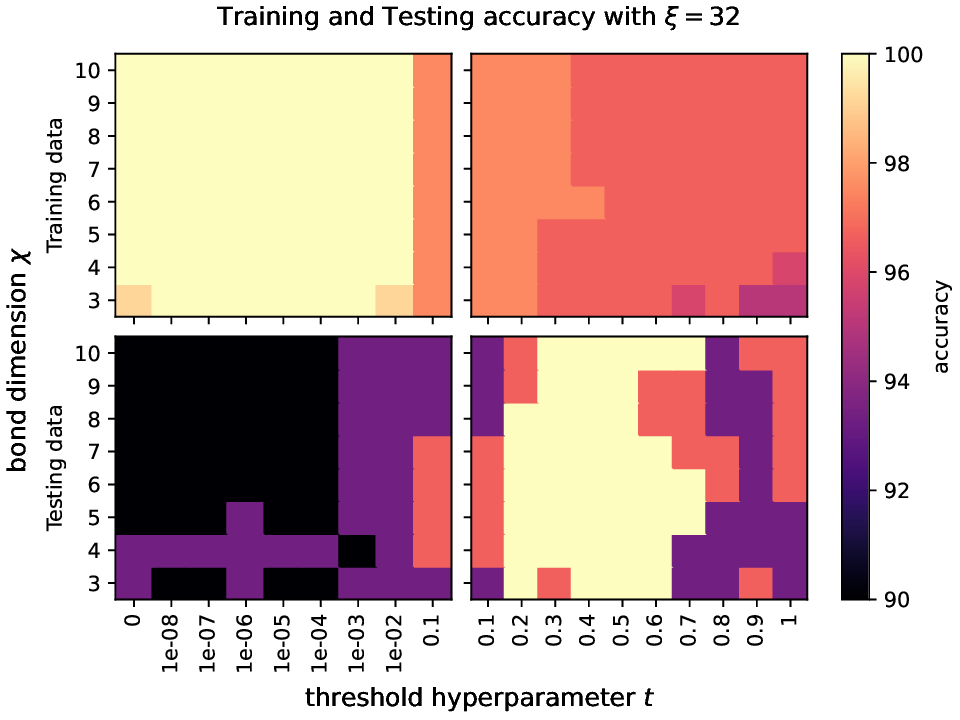}
        \caption{Accuracy for $\xi = 32$}
        \label{fig: xi32 accuracy}
    \end{subfigure}
    \caption{Comparison of $\chi$ and $t$ for the reduction operator dimension $\xi = 2$ and $\xi = 32$. For each subplot the top row is the training data and the bottom row for the testing data. On the left is the logarithmic scale for small thresholds, surveying the different orders of magnitude of the maximally allowed normalization, while on the right is a linear scale that transitions the HTN to a QTN.}
    \label{fig: data 2D}
\end{figure}

For the plots in Fig.~\ref{fig: data 2D} the Iris dataset\citep{Fisher1936,Anderson1936} was used. The HTN in Fig.~\ref{fig: HTN explanation}.2 has the following hyperparameters besides our newly proposed: the bond dimension $\chi$ connecting the isometric tensors, the bond dimension $\xi$ of the reduction operators. If $\xi = 2\chi$, the HTN can express any unnormalized MPS so precisely the architecture pioneered in \citep{Stoudenmire2016}, while setting $\xi = 2$ reduces the number of required ancillas. The small models are trained with 20 two-site sweeps, where at each step of the sweep the local tensor is close to fully optimized using the ADAM optimizer.

For both $\xi = 2$ and $\xi = 32$, the impact of our newly introduced hyperparameter that manages the level of allowed post-selection is larger than that of the bond dimension $\chi$. For the model with $\xi = 32$ that may emulate the respective CTN architecture completely, we observe that the training loss is even invariate with $\chi$. For the training loss, we observe the monotonicity of the loss over the hyperparameter derived in Prop.~\ref{prop: monotonicity}. Thus, the hyperparameter can serve as a natural method to limit the amount of post-selection permitted in a quantum model, while the corresponding loss function allocates post-selection efficiently. For the testing data, we observe overfitting for small $t$, i.e. for a lot of allowed post-selection as can be seen in from the small training loss and large testing loss in Figs.~\ref{fig: xi2 loss} and \ref{fig: xi32 loss}. This overfitting causes a collapse in the accuracy of the obtained models, see Figs.~\ref{fig: xi2 accuracy} and \ref{fig: xi32 accuracy}. The emergence of overfitting at small $t$ is due to the fact that a smaller $t$ corresponds to a larger expressivity of the HTN. The overfitting appears slightly stronger for larger bond dimensions $t$. Furthermore, much of the impact and observed behaviour of post-selection is already obtained for small $\xi = 2$, which makes it practical for QML models as few ancillas are needed.

We probe the model capabilities on a dataset with more features a simple binary classification of the MNIST dataset that is rescaled to $7\times 7$. Although, the model from Fig.~\ref{fig: QML model} can not accurately represent quantum channels for a larger number of sites due to limited bond dimensions. In our test case using $h=1$, i.e. no post-selection, we ran into a barren plateau \citep{McClean2018} with the model returning a fully mixed state for data derived and random initializations. This amounts to random guessing of the output which persists during training due to a vanishing gradient. Additionally to the effects of the curse of dimensionality the machine epsilon increase because of numerical accuracy of the environment caching. For $h=0$ the final testing accuracy was $99.68\%$. The other model settings were chosen to be $\chi = 10$ and $\xi = 40$ for both cases of $h$. Thus, the barren plateau is counteracted by selecting subspaces of the Hilbert space that have non-zero gradients.

\section{Conclusion and Outlook}\label{sec: conclusion}
In this work, we identified the core differences between classical and quantum tensor networks. On the basis of a procedure that allows inference of CTNs on quantum computers, we derived a unified framework that allows for a smooth transition between CTNs and QTNs. Next to the already established bond dimension, we found that post-selection is the property that distinguishes between CTNs and QTNs. We presented the ability of post-selection to improve global separation of data in QML. Because it is a limited resource in real systems, we present a hybrid HTN accompanied by a threshold hyperparameter that allocates a given level of post-selection to the HTN by connecting it to the normalization of the post-selected output.

The numerical results underline that the novel application of post-selection with the presented hyperparameter can benefit QML. Post-selection effectively implements a selective or abstaining classifier by discarding low-confidence samples. However, as a trade-off exists with the number of retained shots. The results are for a specific encoding and it is possible, likely even, that better pre-processing can achieve similar results or even surpass this post-processing routine. Finding a good encoding for QML brings its own set of challenges, while this post-selection based approach can be easily implemented. Because post-selection can detect and remove information that worsens the effectiveness of QML models, the question follows, if post-selection can be used to identify better encodings? To this end, further research is needed to link post- with pre-processing in QML. 

This novel approach leads to several other open questions for our future work. A hyperparameter that connects the HTN to a CTN is missing to fully connect quantum and classical models, although the transition is on the basis of post-selection, one path to such a quantity might be enforcing the purity of the output before post-selection is applied which leads to a CTN. This would deepen the understanding for which encoded data the CTN or QTN achieves better results, which would help with the question for which data QML is advantageous. Additionally, while the hyperparameter already allows the trainable allocation of post-selection, it is important to understand how it gets distributed across the HTN and how that relates to the data and encoding. This would lead to a better understanding of weak points of a data and encoding pairing and thus inform on the choice of encoding.

\section*{Declarations}
\subsubsection*{Author Contributions}
G.J developed the main conceptual ideas, set up, carried out, and analyzed the experiments, and provided the initial draft. H-M.R supervised and guided the project. K.B imparted guidance and contributed to the final draft. M.P provided guidance towards the simplification of the theory laid out in the initial draft.
\subsubsection*{Conflicts of Interest}
The authors declare no competing interests.
\subsubsection*{Data availability}
Data utilised during the current study are available from the corresponding author on reasonable request. The simulations are available from the DLR but restrictions apply to the availability of these and are therefore not publicly available.
\subsubsection*{Funding}
G.J received funding from the DLR (German Aerospace Center) through the Quantum Fellowship Program. Furthermore, the project was enabled by the DLR Quantencomputing-Initiative and the German federal ministry of Research, Technology and Space; \url{https://qci.dlr.de/projekte/qutenet} 

\nocite{Gray2018}
\bibliography{literature}
\appendix
\section{The Hyperparameter and the MSE loss}\label{app: MSE loss proofs}
With the partial normalization $\mathcal{N}_h$ dependent on the hyperparameter $h$ (either $t$ or $w$) the MSE loss is defined as
\begin{equation}
    L_{\mbox{\tiny{MSE,}}h} = \frac{1}{N} \sum_{i=1}^N \frac{1}{2} (\tau_i - \mathcal{N}_h \circ \Lambda[\sigma_i])^2.
\end{equation}
Again, for $h=0$ the normalization is fully employed and thus we find it a suitable loss function for the general HTN architecture. Now we need to show that for $h=1$ the loss function is optimal for only QTNs. To this end, for any HTN with Eq.~\ref{eq: post-selected channel} we may define a channel that uses the post-selection to give a randomized output. For any post-selected output $\rho_{i,D}$ we instead obtain 
\begin{equation}
    \rho_{i,R} = \rho_{i,D} + (1-\tr (\rho_{i,D})) \frac{I}{\dim(\rho_{i,D})}.
\end{equation}
Then
\begin{align}
\begin{split}
    \tr\Bigg((\rho_{i,R} - \tau_i)^2 - (\rho_{i,D} - \tau_i)^2\Bigg) &= - \frac{1-\tr(\rho_{i,D})}{\dim(\rho_{i,D})} < 0.
\end{split}
\end{align}
Thus, following that each individual contribution is negative, we are always able to obtain a QTN that performs equal or better than an HTN. Furthermore, equality is only obtained if no state is post-selected, i.e. $p_i$. This case appears when post-selecting out states that are never occupied. 

Given the strategy of randomizing instead of post-selecting, it may very well be possible to generalize the above proofs for the cross entropy loss and MSE loss to arbitrary loss functions and tie the proof to properties of distinguishability measures \citep{Bengtsson2017} and statements about operator montonicity \citep{Chansangiam2015}. Specifically, using the triangle inequality of distinguishability measures \citep{Bengtsson2017}.
\end{document}